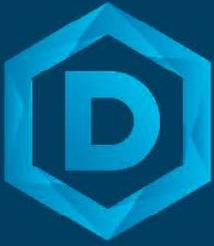

# NATIONAL TREASURE

The Call for e-Democracy and US Election Security

Adam Dorian Wong
adam.wong[at]trojans.dsu.edu
(@MalwareMorghulis)

v1.0

Date: 14 AUG 2024

*"They guard it because they have hope.*

*A faint and fading hope that one day it will flower.*

*That a King will come and this city will be as it once was before it fell into decay.*

*The old wisdom borne out of the West was forsaken.*

*Kings made tombs more splendid than the houses of the living and counted the old names of their descent dearer than the names of their sons.*

*Childless lords sat in aged halls musing on heraldry or in high, cold towers asking questions of the stars.*

*And so the people of Gondor fell into ruin.*

*The Line of Kings failed.*

*The White Tree withered.*

*The rule of Gondor was given over to lesser men."*

*Gandalf on the Decline of Gondor*
*The Lord of the Rings: The Return of the King (2003)*
*paraphrasing JRR Tolkien.*

# Introduction

As the society enters the US General Election, Americans face a grave threat to the democratically-driven Republic. Society can roughly point to the highly-controversial US General Election of 2016 and the attempted takeover of the US Capitol Building on January 6th, 2021. The issue stems from trust. Trust in each other has been strained by incivility of poor leaders. Trust in voting has been warped by influence operations by malicious foreign actors. The nation's greatest asset is the American people. Its greatest national treasure is the institution of free and democratic elections in this Republic. This proposal will not solve influence operations. It works to resolve fragmentation of faith in the elections through providing a solution that can satisfy concerns in a bipartisan way. This is a proposal to safeguard all future elections. The problem with current elections stem from reliance on paper ballots. This is the 21st century, and the US can do better because the technology exists. Identity technologies can be integrated into today's society to improve gaps with REAL ID. Today's voting is chaotic and stressful. Technology can streamline voting and maintain thorough auditable and highly accurate tabulations. A simple view across the Atlantic Ocean, allied countries have succeeded in revolutionizing voting in the digital era. The way forward requires a national smart card identification and Internet-based voting.

# High-Level Arguments Around Election Security

This whitepaper will summarize arguments from the predominantly left (Democratic Party) and predominately right (Republican Party). This dichotomy is examined simply due to their near ubiquitous presence in the US Congress and overall electoral cycle. For sake of civility, one must temporarily ignore counter-intuitive policies and bills that slash election-security funding minimal evidence of election fraud, and attempts to block bills to disclose *dark money* donations [1] [2] [3]. Compromise is the American way in its free elections and an essential style-of-government.

| Democratic Concerns | Republican Concerns |
|---|---|
| Desire for Free Elections | Want Proof-of-Citizenship to Vote [4] |
| Need for Integrity of Voting Systems [5] | Fear of Fraud in Elections [1] |
| Support for Domestic Security | Questioning of CISA's Role [6] |
| Need for Cybersecurity [7] | Cost of Election Security |
| US Leadership in Cyberspace | US is Not Europe |

*Discussion.* Election security is a non-partisan issue. The world looks towards the US as a leader of the free world, with respect to stability and security. Current voter systems are antiquated and the Brennan Center for Justice argues that lower-income Americans are impacted the most because of this ailing infrastructure [8].

# Threats to Domestic Affairs

***Natural Threats to Voting Systems***. Electronics are not fool-proof. Although highly unusual and rare, cosmic radiation can influence data and memory. In 2003, a Single-Event Upset (SEU) occurred when cosmic (gamma) radiation had caused a bit-flip and had added approximately 4,096 extra votes to a candidate in Schaerbeek, Belgium [9] [10]. Radioactive elements are capable of influencing digital systems [10]. Admittedly, the affected system was an e-voting (also called i-voting) system [11]. However, voting systems today incorporate electronics one way or another.

***Internal Threats to Stability***. FBI Director Christopher Wray and DHS Director Alejandro Mayorkas both alluded to an increasing threats facing the homeland [12] [13] [14]. The Nation has a former President claiming election fraud, who endangered the Members of Congress, influenced zealots to overturn the rule of law, and vocally championed insurrection [15] [16] [17] [18] [19]. Special interest groups, unregulated militias, and political bodies may attempt to assert control over the American Republic [20] [21]. This whitepaper will not fix corrupted organizations and individuals [22] [23]. However, the proposed solution through this whitepaper could provide a way forward to having immutable empirical evidence and a higher-confidence towards safeguarding election integrity.

**Information Operations**. Internal and external actors will continue attempts to negatively influence the American Experiment – perhaps to indirectly shape US-policy towards perceived favorable conditions. Misinformation will remain difficult to counter, mitigate, or suppress without strong national and secular education – simply put, a stupid populace is easier to manipulate and redirect resentment of real domestic issues. Adversaries are capitalizing on this fault-line. Vanessa Williamson calls this democratic decline [24]. domestic extremism will be enabled by misinformation such as with Qanon or Project 2025 [25] [26] [27]. The Alt-Right initiative, Project 2025, outlined plans to cripple DHS and CISA – organizations which are responsible for the nation, border security, intelligence dissemination, and election security [24] [27]. Artificial Intelligence (AI) will exacerbate the information-generation and information-diffusion problems [28]. External actors will continue to amplify misinformation to influence domestic affairs and exploit internal conflicts.

To be clear, this proposal will have little-to-no direct effect on misinformation campaigns or the echo-chambers enabled by social media platforms. Simply put, this proposal cannot fix stupidity; a well-educated populace can mitigate the effects of information operations. [29]. Regardless, revolutionizing the methods of voting will provide a better empirical dataset to prove or disprove election integrity to naysayers.

***Counter-Argument to Election Fraud Claims***. A study published through the National Academy of Science demonstrates that claims of voter fraud were "*unlikely or impossible*" [30]. Recent court cases suggest that perpetuation of some individuals had no "good faith basis" to perpetuate lies on election fraud [31]. At the time of this whitepaper, there are approximately 330 million Americans [32]. A subset of this number is eligible to vote and a smaller subset is actually voting. However, not everyone is a criminal or has criminal intent – the number of actual voters will drastically dilute potential instances of fraud.

# Current US Identification

*Identification Cards.* Identification comes in many forms and attributes or collectively categorized as Personally Identifiable Information (PII). Historically, identification cards often pertain to driver licenses and are at the discretion of the individual state, namely their Departments or Registries of Motorized Vehicles (DMV/RMV) [33]. However, these states place their respective DMVs under different departments – some consider Departments of Transportation, others place them under Internal Revenue, and so on [34] [35]. Passports and Passport Cards are a few examples of federal-level identification for private citizens and these typically have a 5-year or 10-year age bracket depending on age bracket [36]. Unfortunately, Social Security Numbers (SSNs) and their cards were never intended to be personal identifiers and private sector began using it heavily in the 1960s with the emergence of computers [37].

*REAL ID Act*. The REAL-ID Act was passed in 2005 following recommendations set by the 9/11 Commission [38] [39]. After several Interim Final Rules and the unexpected COVID19 pandemic, the enforcement of requiring REAL-IDs for access to commercial travel and federal facilities had been delayed to 2025 [38] [40] [41]. Individual States lead the designs of REAL-ID compliant cards and DHS provides guidance for compliance [42]. Based on DHS FAQs, several examples demonstrate that watermarks are unique to each state's theme or regional culture. However, federal guidelines are not only apparent, but common or standardized such as: photo, date-of-birth, name, address [43].

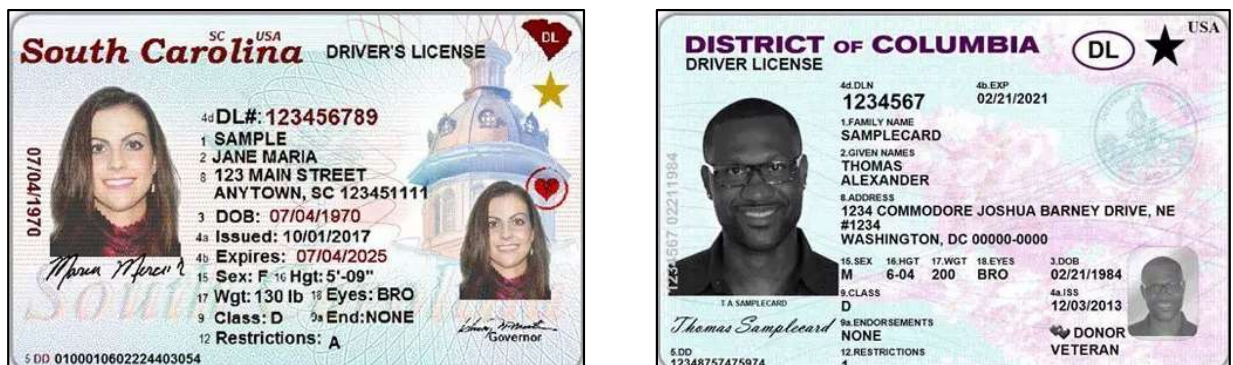

*Figure 1: Examples of READ IDs (Generic Sample) [43]*

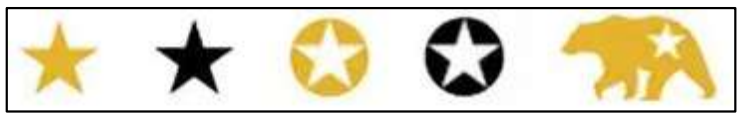

*Figure 2: Example authorized (upper right corner) markings for REAL ID-compliant cards) [43]*

# Public-Key Infrastructure (PKI)

***Public Key Infrastructure (PKI)***. Although, Zero-Trust Architecture (ZTA) is the new hot buzzword, a single word comes to mind: *identity*. Identity Access Management (IAM) is the cornerstone of tracking user activity and actions on a network. Identity is based around Public-Key Infrastructure (PKI). PKI operates off of chains-of-trust with the Root Certificate Authority at the top and subordinate Intermediate Certificate Authorities below validating lower echelons and individual Personal Identity Verification (PIV) cards [44] [45]. PKI often integrates with Microsoft Active Directory and leverages Certificate Revocation Lists (CRL) or Online Certificate Status Protocol (OCSP) for validation of trust chains [46]. Aside from identity, it is also used commonly with code-signing or server validation in secure web traffic [47]. Although, this framework is imperfect and does not completely eliminate bad actors from abusing trust such as the ban of WoSign/StartCom cross-signed certificates or malware authors using accessible CAs like Let's Encrypt [48] [49]. Sure, nothing is trusted, but PKI provides a starting point and point-of-confidence within a system-of-systems through mathematics. A pair of digital keys are generated: public and private [50] [51]. These keys are mathematically related through different algorithms such as RSA and ECC. Common Access Cards (CAC) are PKI-based credentials – both as a physical token and as an authenticator for digital identity [52].

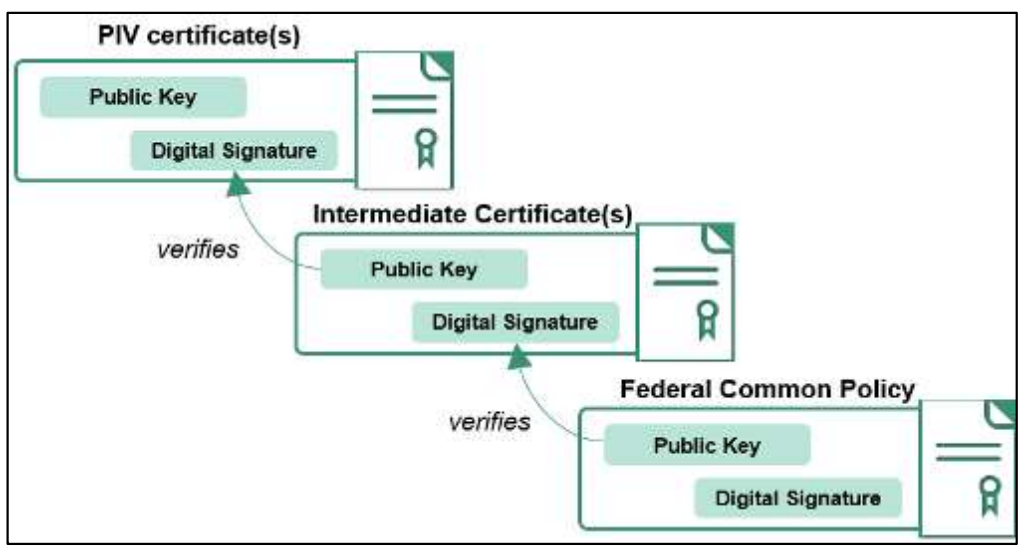

*Figure 3: Example of the PIV PKI Certificate Chain [44]*

***Smart Cards***. The Common Access Card (CAC) is a Personal Identity Verification (PIV) Card commonly found in large commercial enterprise networks and across the US Government, especially within the Department of Defense [53] [54]. Smart cards integrate with Public Key Infrastructure (PKI) and contain certificate chains for card holder identify validation. PIVs are both physical identification and hard-tokens and used in multi-factor authentication within networks [55]. Naturally, the US Office of Personnel Management (OPM) outlines credentialing standards congruent to Homeland Security Presidential Directive-12 (HSPD-12) of 2004 [56]. The directive mandates "*secure and reliable forms of identification*" with resistance to *tampering*, *counterfeiting*, and *fraud* [57]. The execution of the directive is seen in Federal Information Processing Standards Publication (FIPS PUB) 201-3 [58]. The data interfaces for PIVs are outlined in NIST SP 800-83-4 [59]. As source components for identity, biometrics for are outlined in NIST SP 800-76-2 [60]. Additionally, PIVs across the federal government have electronic interoperability and administrative reciprocity between organizations [61] [62].

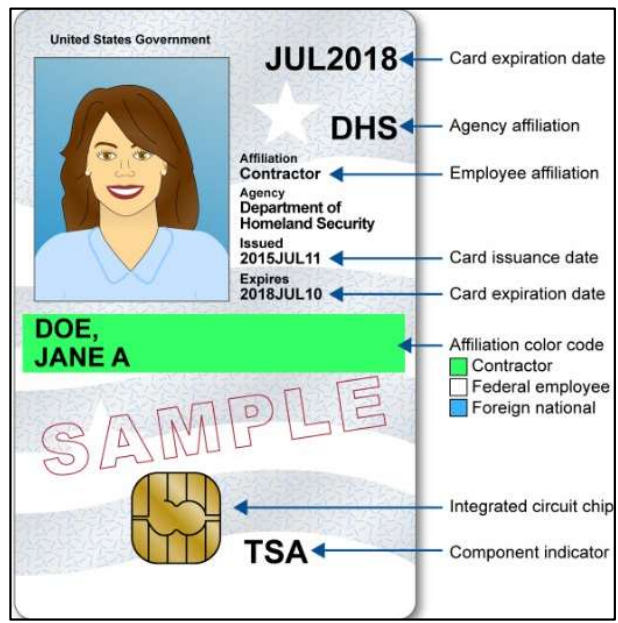

*Figure 4: Example Mockup of Personal Identity Verification (PIV) or Common Access Card (CAC) [55]*

**Mobile Driver's License (mDL)**. The Apple Wallet and Google Wallet have revolutionized the method of paying for goods and services by using a mobile device as the hard-token and payment-enabler. In both instances, data is stored within secured escrows on the mobile devices congruent with ISO/IEC 18013-5:2021 which outlines controls and standards for mobile Driver's License (mDL) [63] [64]. Amazon Web Services (AWS) has private certificate authorities (Cas) capable of providing chains-of-trust for mDL [65]. The following states are part of the Apple Digital License pilot: Arizona, Connecticut, Georgia, Iowa, Kentucky, Maryland, Oklahoma, and Utah, Hawaii, Mississippi, Ohio, and Puerto Rico [66] [67]. The technology appears to be in its infancy and not yet ratified across the entire US. Data appears to be added based on optical character recognition and verification requires users to snap photos of their license and upload corresponding "*selfies*" and head movements, which proceed to the state for verification. This technology relies on Near-Field Communication (NFC) to transmit data and expedite movement through TSA security checkpoints [68]. It can be reasonably assumed that early implementation relied on manual verification. With minimal information on back-end processes, it is currently unknown which manual, automated, or hybrid processes are more effective at intercepting fake physical ID cards or advanced deepfakes.

**Digital Veteran ID Card**. Since September 2022, the US Department of Veteran Affairs (VA) is moving away from provisioning veteran medical cards and towards a digital identification card, given the ubiquity of mobile devices among veterans [69]. Arguably, this could also be seen as both a cost-saving measure and a hygienic decision (during the COVID19 pandemic).

***Login.gov***. The VA has shifted their identity & access management model towards highly-accessible federated infrastructure. For instance, veteran web access is enabled through multiple avenues: login.gov, id.me, DS Logon, and My HealtheVet. Although, the latter two options will be removed by 2025, ID.me accounts allow Single Sign-On (SSO) via common providers like Apple, Google, Facebook [70] [71] [72]. Login.gov is not an identification alone. However, this account access leads to a 1:1 association of person to identity. Login.gov is already used for accessing other government websites. This could potentially be a starting point towards unifying all of the municipal, local, state, territorial, and federal credentialling for the populace.

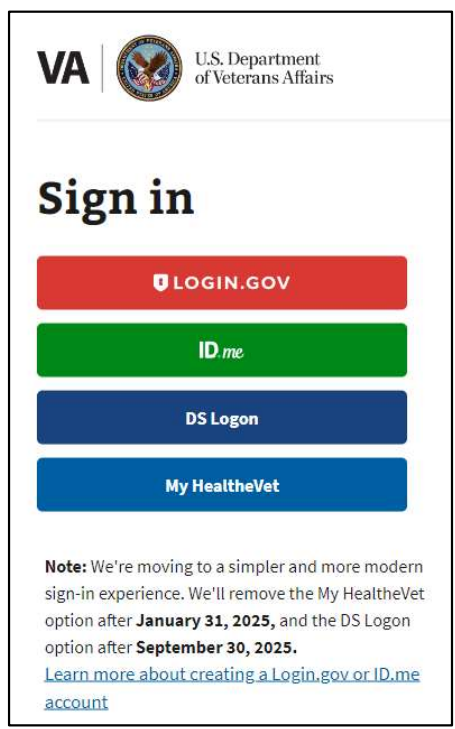

*Figure 5: Example VA.gov login options [69].*

***Discussion***. The problem with digital IDs is that not everyone may be able to afford smartphones. This is not an argument to limit options in novel or revolutionary technologies. Digital IDs may well be the future. However, society cannot forget less-fortunate individuals who may not have access to such luxuries, especially with respect to voting in a free and democratic nation. There may be costs for ID renewal or visiting the DMV, or even up-front fees for an ID. Arguably, the bar-of-entry for a physical ID would be far lower than a smartphone containing a digital ID. However, digital IDs could still be an alternative token in the proposal for election security. Options promote democracy.

# Current Threats to US Voting Practices

      Current paper ballot counting machines are still susceptible to tampering, especially network-connected ones. Some researchers note the technology for ballot counting is still relatively infant and contemporary [73]. There are few if any edge cases where systems were tampered [74]. The US Election Assistance Commission (EAC) outlines security measures for the physical security of hardware and CISA provides support on security efforts [75] [76] [77].

      Current practices for counting ballots are time-consuming, antiquated, and still difficult to audit. Some private industry companies, special interests, and lobbyists may very well economically benefit from refusing evolution and pressuring the US to rely on vestiges of darker ages and long bygone eras for the sake of tradition. Difficulty is measured in time and level of work. Granted, hand-counting is very imperfect, but still a viable audit approach while new technologies are researched. However, America can do better and deserve better.

| Issues | Description |
|---|---|
| Electoral College | • Number of electoral delegates is tied to the total number of members of Congress (across both houses). However, the process waters down the popular vote to 538 electoral votes [78] [79]. It marginalizes and suppresses the will of the American People [80]. This system stays in place because there are currently no better systems for ballot counting. Some candidates have won the Popular Vote, but only to fail and lose an election due to the Electoral Vote [81] [82]. |
| Human Error | • Ballots may be miscounted in highly-contested election cycles which delays assumption-of-leadership and under conditions of fatigue [83] [84] [85] [86] [87].<br><br>• Hand-counting is useful as a secondary audit, but still inaccurate and inefficient [87] [88].<br><br>• Paper ballots and hole-punched voting led to the counting crisis of the US Election of 2000 where "chads" weren't completely disconnected from perforations [89]. |
| Volunteer Hours | • Poll volunteers place numerous hours of work into supporting the democratic process. Their compensation varies if at all [90]. |
| Legacy Technology | - Although now banned, recent elections used antiquated communication hardware or unsecured modems to transmit data [91]. |
| Record Retention | - It is unclear on how long voting ballots are maintained before secure destruction. Hand-marked ballots have been at the nexus of controversy in other state elections [92]. |
| Hardware Accountability | - Hardware is kept in secured locations, but every year, hardware is shuffled around. In 2022, a missing voting machine had been sold on eBay of all places [93]. |
| Voting Convenience | - Federal Law does not specify that private employers must give time off to vote [94].<br>  - Individual states have discretion over enforcement and not all states provide time-off [94] [95].<br>  - The average time-off to vote is about 2 hours and individuals pending traffic, administrative work, and parking even if polls are open all-day [95].<br>- Office of Personnel Management does specify time-off to vote– for federal employees [96].<br>- Absentee ballots are available for remote-voting, but these are prone to delays, administrative errors, or damage [97].<br>  - These have also caused confusion during contested races [98]. |
| Biased Actors or Insider Threats | - Malicious insider actors could attempt to subvert the America's electoral process by infiltrating volunteer groups and tampering with voting systems [74].<br>- Grassroots efforts at weaker echelons of government (tribal, local, municipal, state) are more susceptible to influences of bias, harassment, intimidation, or subversion [99]. |
| Influence Operations | |
| | - Domestic groups have claimed election fraud and propagate conspiracy theories from echo chambers [100].<br><br>- Foreign adversaries capitalize on manipulation, misinformation, and narrative shaping to foster internal dissent, to create discord, and to exacerbate domestic fissures, in order to weaken the target society [101] [102] [103]. |

*This table is not all-inclusive. It merely highlights different threats to US Elections.

***Discussion***. This is insanity that the nation continues to rely on pen-and-paper. There is no drive or incentive to find a bipartisan method to secure elections – provided that the issue gives both sides talking-points to muse and argue over every four years. Network-connected systems are not flawless. America continues to face cyber threats. To the contrary, it is possible to create a resistant and resilient voting system and one that welcomes the modern technological age.

A looming problem is that voting sites do not attribute individual ballots, but rather check if a voter is eligible and registered to vote and if the individual has voted at all. This makes sense because historically, Americans are very private about voting and rely on the secret ballot method. To Americans, this is necessary when passions and prejudices contradict the sentiments of neighbors or peers in positions of authority like landlords [104]. A national PKI-based ID and electronic voting system could fix this.

# Case Study: Estonia

*e-Democracy (i-Voting)*. Estonia is a staunch NATO ally and no novice to the technology age. It is also the pioneer and leader among the NATO consortium through exploring topics of cyberwarfare law via the Tallin Manual [105]. Since 2005, Estonia has relied on e-IDs and Internet Voting (i-Voting). Although, Estonia is a rather small country compared to the US, they have embraced the digital age by integrating it with healthcare, public services, auditing with blockchain. Digital attacks still occur with i-Voting systems, especially to technologically-advanced countries like Estonia. However, it has developed a robust voting process and cyber resilience to cyberattacks [106]. The movement to i-Voting requires commitment to its security (secure by-design, resources, etc.), identity verification, and a level of faith in technology (ease-of-use, adaptation by the populace) [107]. Voting requires digital signatures through ID cards, smart card readers, and PINs [108]. Votes can change or happen multiple times a day, but repeated votes are removed [109]. It can be reasonably inferred that Estonia probably uses a variation of blockchain to track multiple votes from an individual to account for flip-flopping or accidental votes. Additionally, Estonia solved the anonymity issue with i-Voting by redacting information from reaching the auditors at the National Electoral Commission [110]. Interestingly, the voting period is 1 week long, uses the last ballot cast as the counted vote, and authenticated users have the option to vote anywhere using their smartcard national ID or their mobile e-ID (their mDL) [111]. Researchers applauded this revolutionary approach to voting, albeit, the system is also imperfect as insider threats are still a concern on the server-side [112]. This voting system spawned from influence operations in the early 2000s and is a target of criticism [113]. Regardless, this e-Democracy or i-Voting approach has solved accessibility, security, and accuracy of client-side digital voting.

*Discussion*. Some people would argue that the US is too large for this sort of complex infrastructure. This country has done magnificent things over the course of history. America has not tried in this endeavor. The Nation will not know i-voting will fail unless it sincerely tries first. America has an opportunity to make voting tabulations empirical and precise. Sure, goofy individuals may try to flip-flop votes for fun – but a cutoff time to freeze votes with last-vote counts rules will mitigate this issue. This infrastructure will require commitment to its security and resources through the help of USCYBERCOM, DHS, and CISA. America must work smarter, not harder. Naturally, this could work with PKI-based national ID cards because every young adult receives a driver's license. America needs to stop worrying and learn to live with the digital bomb. Attacks will happen physically or digitally. However, the US must adapt to the times and explore new technology.

# A Proposed Future of Election Security

It is recommended that the US move towards a national PKI-based identification system and support true i-Voting to simplify and secure its elections. This proposal is not all-inclusive or by any means perfect. This whitepaper is proposing an evolution to US elections and implementation of REAL ID 3.0 with these high-level guidelines:

**Implementation the evolution of the REAL ID to version 3.0.**

1. Federal Government:

    a. Would enforce use of a separate set of PIV cards for private citizens.

    b. The Federal Government would be responsible for the PKI Root Certificate Authorities within GovCloud.

        i. Specifically, DHS and CISA would be responsible for securing PKI cloud instances -- given DHS lead in customs & border security, cyber infrastructure security, and responsibility over *.gov websites.

        ii. Department of Homeland Security, Department of Treasury, Internal Revenue Service (IRS), and Social Security Administration (SSA) would collaborate or be consulted on identity verification.

        iii. OPM would be consulted on how to implement similar, but not identical infrastructure at a national-scale.

2. Individual States:

    a. Would act as Intermediate Certificate Authorities in the certificate chain.

    b. DMVs and RMVs will maintain their ability to generate PIV cards.

    c. Would be responsible for provisioning certificate-loaded PIV cards to the populace, congruent to law and regulation.

    d. Would maintain authority to add their own unique watermarks or flair, provided it is congruent to requirements outlined in the REAL ID Act.

    e. IT Departments would be responsible for the infrastructure, cyber compliance, and security of their individual tenants.

    f. Similar documentation requirements would apply to REAL ID 3.0.

i. Unique identifiers (UIDs) would link to Social Security Numbers (SSNs), although it not used for financial purposes and UIDs would be align to a federal standard.

ii. SSNs would not be stored on the PIV, but associated in an encrypted back-end database.

g. National Guard Cyber Mission Forces would be authorized to conduct Defensive Cyber Operations – Internal Defense Measures (DCO-IDM) while under Title 32 activations to protect & monitor this PIV infrastructure.

3. PIV cards:

a. Acts as a normal physical ID or driver's license for normal day-to-day use.

b. As previously stated, would have a certificate chain associated to an individual citizen.

c. Would be highly-accessible as both digital ID and physical ID.

d. Usable identification and authentication for i-voting – anywhere and at any time.

i. NFC-enabled & PIN

ii. Chip & PIN

e. Provides assurance that final votes happen once and in one place.

4. eDemocracy or i-Voting:

a. Registration and votes occur anywhere, any time by widening the window of vote time for sound decision-making.

b. Orchestrated through a secure government cloud portal.

c. Enables faster and efficient voting.

d. Voting can flip or change within the appropriate safeguards, defensive throttling analytics, and / or time window, until finalized by the hard-stop poll-close deadline.

i. States could define their own safeguards (congruent to their political climates). However, the States must provide the clear

        public awareness training on what is appropriate or what safeguards exist for vote-flipping.

e. Auditable, digitally and physically – traceable to an individual via blockchain.

    i. Sample audits can be sent to individuals to verify identity or validate voting activity.

f. Volunteers can still work at designated sites to teach individuals how to e-vote, how to e-register, validate voting actions, review voting results, or conduct overall voting audits.

# Conclusion

  Although, this proposal is merely an idea and is imperfect. There are complex risks involved – where to start, how to start, how to minimize cost, how to maximize effectiveness, and so on. The alternative is that the country does nothing. There is risk in everything Americans do and the way Americans live. With information operations by foreign influence and exacerbation by bad domestic actors, the American people lack trust in each other during election cycles. Why not place our trust in empirical values – in numbers. People trust metrics. America must let data drive our decision-making. A national PKI-based identity smartcard (with digital mobile ID) and e-Democracy technologies are the way forward to securing all future elections. Sure, the current voting system will work just fine until it doesn't. The American Experiment has successfully maintained a democratically-elected republic since its founding. However, the country is still constrained by vanity of archaic traditions while adversarial attacks advance in complexity. Weaker administrations emboldened our adversaries, authoritarians, autocrats, and tyrants - through words of admiration rather than condemnation. America must maintain its leadership across all domains and especially with respect to technology in election systems. Technology will never make problems disappear, but technology will make problems smaller and more manageable. Computers work faster than humans. If trust in computers is in question, then America needs to work harder in DevSecOps and in secure-by-design software. After two-and-a-half centuries, American election methods must evolve with the society. This eDemocracy (i-Voting) has the potential to work and change America for the better. Unless, of course, both sides of the aisle merely wish to maintain an illusion of choice and to continuously enable the vicious cycles of politically-convenient strawman arguments through paper ballots excuses.